\documentclass[journal]{IEEEtran}

\usepackage{amsmath,graphicx}
\usepackage{algorithmicx}
\usepackage{algpseudocode}
\usepackage{algorithm}
\usepackage{url}
\usepackage{multirow}
\usepackage{booktabs}
\usepackage{xcolor}

\usepackage{amsmath,amsfonts,amssymb}

\ifCLASSOPTIONcompsoc
 \usepackage[caption=false,font=normalsize,labelfont=sf,textfont=sf]{subfig}
\else
 \usepackage[caption=false,font=footnotesize]{subfig}
\fi

\begin{document}
\title{Wide Color Gamut Image Content Characterization: Method, Evaluation, and Applications}

\author{Junghyuk~Lee,
        Toinon~Vigier,
        Patrick~Le~Callet,~\IEEEmembership{Fellow,~IEEE,}
        and~Jong-Seok~Lee,~\IEEEmembership{Senior~Member,~IEEE}%
\thanks{
This research was supported by the MSIT (Ministry of Science and ICT), Korea, under the ``ICT Consilience Creative Program'' (IITP-2019-2017-0-01015) supervised by the IITP (Institute for Information \& communications Technology  Promotion), the International Research \& Development Program of the National Research Foundation of Korea (NRF) funded by the Korea government (MSIT) (NRF-2016K1A3A1A21005710), and the Science and Technology Amicable Research (STAR) Program funded by the Partenariats Hubert Curien (PHC) and the Campus France (PHC-STAR 36664WK).}
\thanks{A preliminary version of this work was presented at the International Conference on Image Processing (ICIP) in 2018~\cite{ICIP2018}.}
\thanks{\textsuperscript{\textcopyright} 2021 IEEE. Personal use of this material is permitted. Permission from IEEE must be obtained for all other uses, in any current or future media, including reprinting/republishing this material for advertising or promotional purposes, creating new collective works, for resale or redistribution to servers or lists, or reuse of any copyrighted component of this work in other works.}
}


\maketitle

\begin{abstract}
In this paper, we propose a novel framework to characterize a wide color gamut image content based on perceived quality due to the processes that change color gamut, and demonstrate two practical use cases where the framework can be applied.
We first introduce the main framework and implementation details.
Then, we provide analysis for understanding of existing wide color gamut datasets with quantitative characterization criteria on their characteristics, where four criteria, i.e., coverage, total coverage, uniformity, and total uniformity, are proposed. 
Finally, the framework is applied to content selection in a gamut mapping evaluation scenario in order to enhance reliability and robustness of the evaluation results.
As a result, the framework fulfils content characterization for studies where quality of experience of wide color gamut stimuli is involved.

\end{abstract}


\section{Introduction} \label{sec:introduction}
\begin{figure}[!t]
\centering
\includegraphics[width=3.5in]{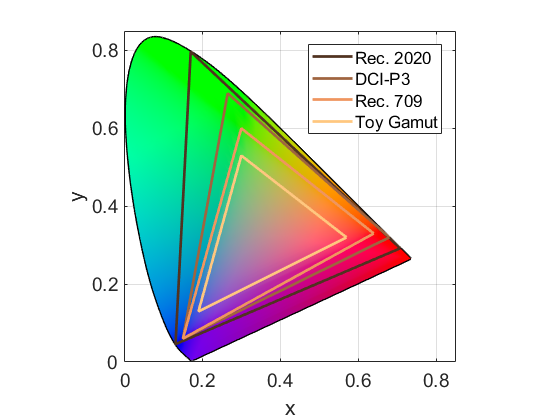}
\caption{Color spaces of color gamuts considered in this work on CIE 1931 chromaticity diagram.}
\label{fig_testgamut}
\end{figure}

\IEEEPARstart{I}{n} order to provide more realistic and higher visual quality of experience (QoE) of multimedia contents to viewers, technologies related to wide color gamut (WCG) have emerged. Since the HDTV standard ITU-R Rec.709~\cite{ITU709}, several WCGs have been proposed. International Telecommunication Union (ITU) approved Rec.2020~\cite{ITU2012} as the standard color gamut for UHDTV, which covers the widest area of the CIE 1931 space~\cite{CIE1931} (see \figurename~\ref{fig_testgamut}). Recently, many devices including mobile devices support WCGs as a process of transition to Rec.2020~\cite{UHDTV}. Considering various environments of multimedia content consumption, gamut mapping is often inevitable in order to match the original color to displaying devices.

In this situation, several gamut mapping algorithms (GMAs) have been proposed as well as the standard algorithms in the CIE guideline~\cite{CIEguidelines}. Among them, gamut reduction aims to reproduce details and color quality of WCG images in smaller gamuts, and maps colors from a large source gamut to a smaller target gamut. For instance, a gamut reduction algorithm proposed in~\cite{GRA} iteratively modifies the color of each pixel based on adaptive local contrast according to the Retinex theory~\cite{Retinex}. In developing and evaluating methods related to color representation of visual contents including WCG and GMA, it is important to assess how the result will be perceived by human observers. In order to assess perceived QoE, subjective and/or objective studies are usually conducted~\cite{Saliency,Blind2016,Sampling,Blind2018,IQA2018}. 

When subjective or objective QoE evaluation is conducted, one of the primary steps is to select a representative and compact set of source contents, whose processed versions are assessed. This step, equipped with a proper content characterization method, is important not only to conduct an experiment efficiently with limited resources (especially for subjective evaluation) but also to draw reliable and reproducible conclusions. If the contents for an experiment are biased and not representative in their characteristics, the results may be biased and not be generalizable for other types of contents. Thus, it is important to select representative contents according to the purpose of a specific experiment. Towards this, it is necessary to objectively measure the representativeness and suitability of a set of contents.

In this paper, we propose a novel framework to characterize WCG contents and its applications.\footnote{Our code is publicly available at https://github.com/junghyuk-lee/WCG-content-characterization} We note that WCG contents are frequently exposed to the gamut mapping processes targeting diverse displaying environments. Therefore, it is important to consider the perceptual difference caused by gamut reduction for the WCG contents. Thus, our main idea is to measure perceptual difference due to successive gamut reduction in order to characterize a WCG content. We also validate the framework by applying it to two applications involving content characterization and selection in practical WCG-related studies.

Our main contributions are summarized as follows:
\begin{enumerate}
\item We propose an objective framework for WCG content characterization based on perceptual properties related to differences due to color gamut change.
We obtain the perceptual difference due to gamut reduction by predicting the subjective score with an objective metric.
\item In order to demonstrate its effectiveness, we apply the framework to practical applications related to WCG.
As one of the applications, we propose multiple criteria characterizing WCG datasets quantitatively based on the proposed perceptual difference. 
Using them, we conduct analysis of existing WCG datasets.
\item In addition, we apply the framework in a scenario of benchmarking GMAs.
We demonstrate that the reliability of the benchmarking is maximized by content selection using the proposed framework.
\end{enumerate}
Note that this paper has distinguished contributions compared to our preliminary work~\cite{ICIP2018} in various respects. While the preliminary work only introduces the basic idea of the proposed framework, this paper provides its detailed description and further analysis along with the shared source code. In addition, we present the two practical applications involving WCG contents and demonstrate the effectiveness of our framework for content characterization.

The rest of this paper is organized as follows.
In Section~\ref{sec:related}, we briefly survey the related works.
In Section~\ref{sec:method}, we present the proposed framework for WCG content characterization and provides its implementation details.
In Section~\ref{sec:DBA}, we describe how the proposed framework is applied to quantify characteristics of WCG datasets and provide analysis of existing WCG datasets.
In Section~\ref{sec:scenario}, we describe another use case of the proposed framework for comparison of GMAs.
Finally, Section~\ref{sec:conclusion} provides concluding remarks.

\section{Related Works} \label{sec:related}
\subsection{Gamut Mapping}
In order to reproduce the original color of contents in devices having smaller color gamuts, several GMAs have been proposed. 
They can be categorized into global and local strategies. 
The former changes all colors of out-of-gamut pixels towards the inside of the target gamut by gamut compression or clipping~\cite{GMA1988, GMA1989,GMA1997,GMA1997M, GMA1999}. 
QoE of the gamut-reduced image often decreases since the color may become blurred around the pixels where the color changes. The latter considers spatial relationship between pixels at the expense of increased computational complexity in order to enhance perceived quality of the gamut-reduced images~\cite{GRA,spatial2003,spatial2006,spatial2007, Retaining2007, Efficient2007, Spatial2012,Gamut2017}. 

\subsection{QoE Assessment of Gamut Mapping}
QoE of gamut-mapped contents is usually assessed by conducting subjective or objective studies. In~\cite{Perceptual2008}, a psychophysical experiment is conducted to evaluate four GMAs, where the subjective quality of the gamut-reduced images is assessed. In~\cite{CIELAB,SCIELAB, iCAM, hueangle, UIQ}, various color image difference metrics are proposed to measure objective quality of gamut-mapped images. In~\cite{Evaluation2006}, subjective scores of gamut-reduced images using different GMAs are obtained by a psychophysical experiment, and are used to evaluate four objective metrics. 

However, in~\cite{Evaluating2008}, it is concluded that the color difference measured by objective metrics and the perceived image difference between original and gamut-reduced images do not correlated well. There are attempts to improve objective metrics by employing spatial filtering that simulates the human visual system~\cite{SHAME} and by extracting features based on perceptually important distortion~\cite{CID}. On the other hand, studies that consider measuring QoE of WCG contents are rare.
In~\cite{Physiological2016}, a physiological experiment is conducted to measure electroencephalography during watching WCG video contents.

\subsection{Content Characterization}
Winkler~\cite{Winkler} quantifies the characteristics of the contents in existing image and video datasets, including spatial information and colorfulness for color images, and motion vectors for video contents, based on which the representativeness of a set of contents can be evaluated~\cite{BVI,MM2019}. In~\cite{Selecting2013}, it is suggested to consider attributes of the test material such as brightness, colorfulness, amount of motion, scene cuts, types of the content, etc. for subjective video quality assessment. In~\cite{Krasula_Features}, contrast, colorfulness, and naturalness are considered to characterize tone-mapped images for HDR contents. In~\cite{LF2017}, a content selection procedure for light field images is proposed using high-level features consisting of depth properties, disparity range of pixels, refocusing features, etc. as well as general image quality features. 

In~\cite{Narwaria_QoMEX2014}, however, it is argued that those simple characteristics do not sufficiently cover the perceptual aspects of visual contents when processing steps (i.e. tone-mapping) are involved. Therefore, an approach is proposed to characterize HDR contents in the viewpoint of whether an HDR content is challenging for tone mapping operators. It focuses on the perceptual change due to the dynamic range reduction that is frequently applied to HDR contents. Using this characterization method, a framework to build a representative HDR dataset is proposed in~\cite{Krasula_JSTSP}. In a similar spirit, we propose a novel characterization framework for WCG contents.

\section{Proposed Framework} \label{sec:method}
\subsection{General Algorithm}
We propose a framework for WCG content characterization based on the perceptual change caused by gamut mapping.
We define WCG content characteristics as degrees of the perceptual differences due to successive gamut reduction.
The overall procedure of the proposed method is summarized in Algorithm~\ref{alg:GeneralAlgo}.

\begin{algorithm}
\begin{algorithmic}
\State \# \textbf{Input} $I_{0}$: WCG source image %
\State \# \textbf{Output} $D$: vector of perceptual differences of $I_0$
\State \# $N$: number of target gamut spaces for gamut reduction
\State \# $G_0$: reference gamut space that covers all colors of $I_0$
\State \# $G_n$: $n$-th gamut space smaller than $G_{n-1}$ ($n=1, \cdots, N$)
\State \# $f_{GR}(I,G)$: function that generates a gamut-reduced image with all colors in gamut $G$ from image $I$
\State \# $PD(I,I')$: function that measures the perceptual difference between images $I$ and $I'$ %
\vspace{0.25cm}
	\vspace{1mm}
    \For{$n = 1:N$}
        \State Generate $I_n = f_{GR}(I_{n-1},G_n)$
        \State Calculate $d_{n} = PD(I_n,I_0)$
        \EndFor		
	\State Obtain $D = {[d_{1}, d_{2}, \cdots, d_{N}]}^\intercal$
\end{algorithmic}
\caption{General framework for WCG content characterization}
\label{alg:GeneralAlgo}
\end{algorithm}

The framework in Algorithm~\ref{alg:GeneralAlgo} produces an $N$-dimensional feature vector of perceptual difference for each WCG source content.
First, we obtain $N$ gamut-reduced images by applying a gamut reduction operator that converts the color gamut of the reference image $G_0$ into a target gamut $G_n$ ($n=1, \cdots, N$).
For each gamut-reduced image $I_n$, we apply an objective metric that measures the perceptual difference from the reference image $I_0$.
Finally, we obtain a feature vector $D$ describing the behavior of the WCG content in terms of perceptual difference due to gamut reduction. 
We can utilize this feature in various applications such as WCG dataset analysis, content clustering, and selection, which will be presented in Sections~\ref{sec:DBA} and \ref{sec:scenario}.

\subsection{Obtaining Ground Truth of Perceptual Difference}
Hereafter, we provide implementation details of the proposed framework.
In Algorithm~\ref{alg:GeneralAlgo}, we use an objective metric $PD$ to measure the perceptual difference due to gamut reduction.
Although various image quality metrics have been proposed in literature, metrics specifically designed to measure perceptual difference of images exposed to color gamut change do not exist. 
Therefore, we conduct a subjective test in which the mean opinion score (MOS) of the perceptual difference between gamut-mapped images is measured.
MOS is then used to benchmark existing color metrics and optimize the best one via nonlinear transformation.

\subsubsection{Data} \label{sec:data}
We collect 54 images consisting of scenes from HdM-HDR-2014~\cite{HdM} and Arri Alexa sample footage\footnote{https://www.arri.com/en/learn-help/learn-help-camera-system/camera-sample-footage}, in short, HdM and Arri, respectively.
HdM contains videos filmed in a professional cinematography environment with dynamic ranges up to 18 stops and a color gamut close to Rec.2020.
Especially, it focuses on the WCG by containing videos with highly saturated color and lights.
The perceptual difference of the videos is large when the gamut is reduced.
Arri is a video sample footage provided by the ARRI company.
Contents of the dataset are in various natural topics with up to the Rec.2020 color gamut.
Compared to HdM, color differences are not large when the gamut is not much reduced.
The collected image set is divided into training and validation sets of 30 and 24 images, respectively.
The HDR images from HdM are converted to the standard dynamic range with a fixed value of exposure.

We use DCI-P3 as a reference WCG, which originates from the cinema industry.
And, we use two target gamuts for gamut reduction (i.e., $G_1$ and $G_2$): Rec.709 and Toy.
With widespread displays abiding by the HDTV standard, gamut reduction from P3 to Rec.709 frequently happens to WCG contents.
In addition, to cover a high degree of gamut reduction, we employ an artificially created gamut, called Toy, which has been used in the state-of-the-art WCG studies~\cite{GRA,gamutex2017}.
It is smaller than Rec.709 and produces large perceptual difference when the gamut of a WCG image is reduced to it.
The choice of these two gamuts is based on our preliminary experiments, where for the gamuts between P3 and Rec.709, the gamut-reduced images are not visually distinguishable from those in P3 nor Rec.709; in addition, gamuts smaller than Toy give rise to too much color distortion in the gamut-reduced images and thus are not practically meaningful. 
For gamut reduction, we consider a simple gamut mapping algorithm because complex and time-consuming algorithms are not preferred in the content characterization process.
Hence, we use the gamut clipping method that maps colors outside the target gamut at the nearest boundary of the target gamut.

\begin{table}[!t]
\renewcommand{\arraystretch}{1.3}
\caption{RGB Primary Colors in the CIE 1931 Color Space}
\label{table_primaries}
\centering
\scalebox{0.95}{
\begin{tabular}{lcccccccc}
\toprule
\multirow{2}{*}{\textbf{Gamuts}} & \multicolumn{2}{c}{\textbf{Red Primaries}} & \phantom{} & \multicolumn{2}{c}{\textbf{Green Primaries}} & \phantom{} & \multicolumn{2}{c}{\textbf{Blue Primaries}} \\
\cmidrule{2-3} \cmidrule{5-6} \cmidrule{8-9}
&\textbf{x}&\textbf{y}&&\textbf{x}&\textbf{y}&&\textbf{x}&\textbf{y}\\
\midrule
DCI-P3 & 0.680 & 0.320 && 0.265 & 0.690 && 0.150 & 0.060\\
Rec.709 & 0.640 & 0.330 && 0.300 & 0.600 && 0.150 & 0.060\\
Toy & 0.570 & 0.320 && 0.300 & 0.530 && 0.190 & 0.130\\
\bottomrule
\end{tabular}
}
\end{table}

\subsubsection{Subjective Test} \label{sec:experiment}

We adopt the paired comparison test methodology~\cite{paired2014} for the subjective test, because the difference due to gamut reduction is mostly subtle perceptual difference rather than large quality distortion.
The reference image in the P3 gamut and one of the gamut-reduced images produced in Section~\ref{sec:data} are shown in a side-by-side manner.
The images are compared in terms of color difference on a three-point scale: no difference (0), slight difference (1), and clear difference (2).

The test is conducted under the standardized test room condition complying with the laboratory condition described in ITU-R BT.500 such as luminance of the monitor, room illumination, observers, etc.~\cite{ITU500}.
We use an EZIO ColorEdge monitor that can display up to the P3 color gamut.
We heuristically crop each image in half-width ($960\times1080$ pixels) to show both images side-by-side on a single monitor.
Participants are 51 healthy non-expert volunteer subjects consisting of 26 males and 25 females, who are screened by a color and vision test.
We obtain the MOS for each of the 60 images (30 source images $\times$ two target gamuts) by taking the average value of the ratings over the subjects.

The test consists of an exercise and a test sessions.
During the exercise session, the test methodology is described to the subjects with five exercise stimuli that are different from the test stimuli.
The test session proceeds sequentially for each pair of images as follows.
First, a reference image and one of its gamut-reduced versions are displayed on the monitor up to five seconds.
Then, the monitor turns into a gray screen.
At anytime during these steps, the subjects can enter their rating using a keyboard.
Finally, the monitor turns into (or stays) gray for one second for a break and then the next pair is shown.
The viewing order of the stimuli is set random for each subject.
The arrangement of the reference image or the gamut-reduced image (i.e., left or right side) is also randomized for each pair.
At the beginning of the test session, three dummy pairs are shown for stabilization, which are also different from the test stimuli.

\subsection{Fitting Objective Metric} \label{sec:fit}

In order to approximate the subjective score of the color difference due to gamut reduction in an objective manner, we employ the color extension of the structural similarity index ($cssim$)~\cite{CD2016,cssim2003}, which can effectively measure perceptually significant structural differences due to gamut reduction between two color images.
The preliminary study~\cite{ICIP2018} shows that it performs best with the highest accuracy among eight commonly-used objective color difference metrics~\cite{CIEDE2000,CIEDES,Colorfulness,ChromaEx,ColorAppear,JCND,AdaptiveSpatio}. 

\begin{figure}[!t]
\centering
\includegraphics[width=3in]{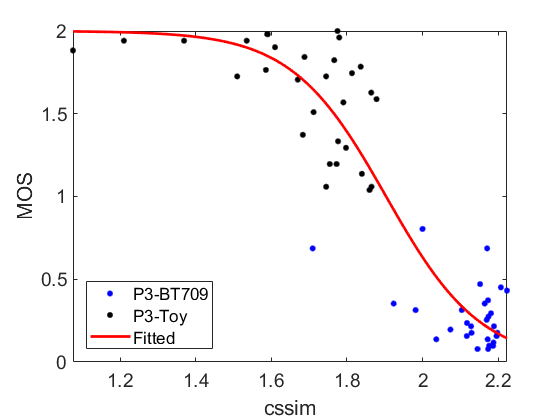}
\caption{Fitted sigmoid function to predict MOS of perceptual difference using $cssim$.
The ground truth and predicted MOS are shown as dots and a red line, respectively.}
\label{fig_Fitcssim}
\end{figure}

For each pair of the reference and gamut-reduced image, we measure the $cssim$ score.
The score is further fitted to the MOS by a monotonic nonlinear function as described in~\cite{ITU149}:
\begin{equation}
f(x) = \frac{\alpha}{1+10^{\beta(\gamma-x)}},
\end{equation}
where the fitted values of the parameters are $\alpha=2$, $\beta=-3.5$, and $\gamma=1.9$.
The result of fitting for the training dataset is shown in \figurename~\ref{fig_Fitcssim}.
In order to evaluate the prediction performance, we obtain MOS for the validation dataset from 20 subjects by following the same procedure described in Section~\ref{sec:experiment}.
The Pearson correlation coefficients (PCCs) between the ground truth MOS and predicted MOS using the fitting function are 0.92 and 0.80 for the training and validation sets, respectively.
Therefore, we calculate the perceptual difference in Algorithm~\ref{alg:GeneralAlgo} as 
\begin{equation}
d_n=PD(I_n,I_0)=f(cssim(I_n,I_0)).
\end{equation}

\subsection{Validation}

We validate the framework by applying it to a simple content selection task.
As mentioned in Section~\ref{sec:introduction}, using representative contents is crucial to draw reliable conclusion in studies on QoE of WCG images.
In the task, the main objective is to select representative images that have diverse behaviors in terms of the perceptual difference due to successive gamut reduction.
We use the framework to obtain predicted perceptual differences due to gamut reduction to the two target gamuts (Rec.709 and Toy) as two-dimensional features characterizing the 24 candidate images in the validation dataset. 
Then, the $k$-means clustering algorithm is applied to the predicted perceptual differences.
The value of $k$ determines the number of representative clusters for content selection, which should be chosen by the user according to the purpose of content selection. In this experiment, we set the value of $k$ to five based on the distribution of the images in terms of the predicted perceptual differences.
One image for each cluster is randomly selected to construct a representative image set, which maximizes the coverage of the feature space.
For comparison, we also apply a random selection method where five images are selected randomly in the same dataset.

\begin{figure*}[!t]
\centering
\subfloat[]{\includegraphics[width=3in]{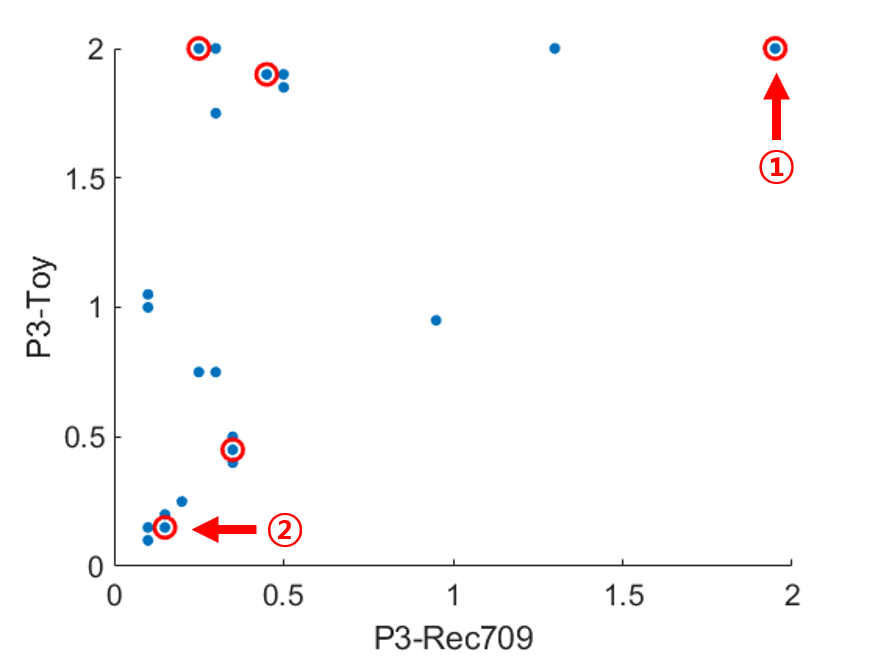}%
\label{fig_selected_f}}
\hfil
\subfloat[]{\includegraphics[width=3in]{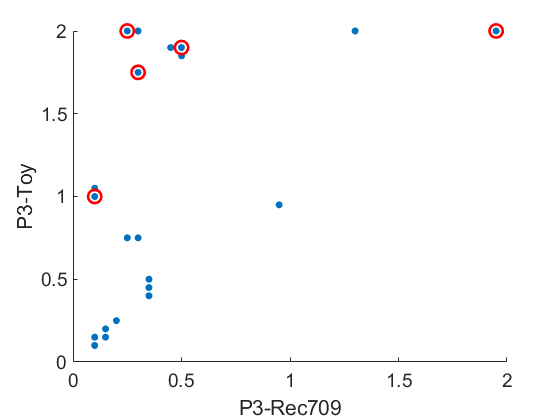}%
\label{fig_selected_r}}
\caption{Example of content selection with (a) our framework and (b) random selection.
The x- and y-axis are predicted perceptual differences of images from the P3 to Rec.709 and Toy gamuts, respectively.
Among the data points shown as blue dots, the selected images are marked with red circles.
Note that the lower-right area of each figure is empty because as the gamut is reduced more, the perceptual difference becomes larger, thus the value of the y-axis would be always bigger than that of the x-axis.
}
\label{fig_selected}
\end{figure*}

\begin{figure*}[!t]
\centering
\subfloat[]{\includegraphics[width=5.5in]{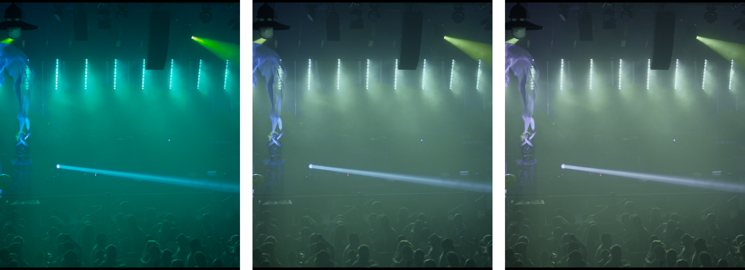}%
\label{fig_selected_large}}
\hfil
\subfloat[]{\includegraphics[width=5.5in]{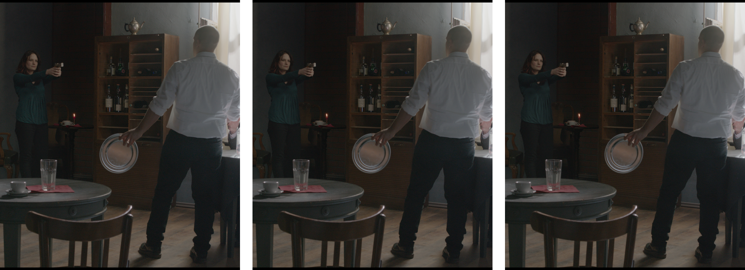}%
\label{fig_selected_small}}
\caption{Selected images corresponding (a) {\large \textcircled{\small 1}} and (b) {\large \textcircled{\small 2}} of \figurename~\ref{fig_selected_f} in the P3, Rec.709, and Toy gamuts (left, middle, and right panels, respectively).}
\label{fig_selected_images}
\end{figure*}

The result for each selection method is shown in \figurename~\ref{fig_selected}.
It can be seen that the selected images by our framework in \figurename~\ref{fig_selected_f} are more spread than the randomly selected images.
In \figurename~\ref{fig_selected_r}, however, the selected images are biased to the upper-side of the feature space.
In this case, images having small perceptual difference by severe gamut reduction are not considered, and the obtained image set cannot be said to be representative.
\figurename~\ref{fig_selected_images} shows two example images (marked in \figurename~\ref{fig_selected_f}) in different gamuts.
In \figurename~\ref{fig_selected_large}, as predicted, large perceptual differences are observed for both gamut-reduced images compared to the reference P3 image, i.e., the overall color of the scene and the green laser lights at the top area.
On the contrary, \figurename~\ref{fig_selected_small} hardly shows any difference between gamut-reduced ones, which is also predicted in \figurename~\ref{fig_selected_f}.
By selecting images with diverse characteristics, a representative dataset can be constructed by our framework.

We also evaluate robustness of content selection with our framework. 
For each of the two methods (random selection and our framework), the selection task is repeated two times to obtain two sets of selected images, and the PCC between the MOSs of the two sets is measured.
We consider that a high value of PCC by a selection method represents a high level of robustness of the method, because it means that the characteristics of the selected images are consistent regardless of repetition or random effects.
We repeat the procedure 100 times.
Much higher PCC values are obtained by our framework than random selection (0.83 vs. 0.15 on average), which is found to be statistically significant via a t-test, $t(137.1)=21.1$, $p<0.001$\footnote{The statistical significance of higher PCC values by our framework is obtained in all cases with $k$ from 2 to 10.}.

\section{Application to WCG Dataset Characterization} \label{sec:DBA}
In this section, we apply the proposed framework to characterization of WCG image datasets.
We describe dataset characterization criteria and analyze existing WCG datasets based on them.
Characterizing datasets helps an experimenter to determine or construct a suitable dataset for studies related to QoE of WCG contents.

\subsection{Dataset Characterization Criteria}
By extending the dataset characterization criteria presented in~\cite{Winkler}, we propose to measure four statistics of perceptual difference measured by the framework as follows.
In~\cite{Winkler}, three statistics of various characteristics extracted from images or videos in the dataset are proposed.
They are two criteria measuring the coverage and uniformity in each dimension, and a multidimensional coverage criterion. 
In addition to these, we also consider the multidimensional uniformity.
Note that we normalize the perceptual differences in each dimension to scale the span of the criteria within [0, 1], i.e., $\tilde{d_i} = d_i/s_i$, where $s_i$ is a normalization factor that is equal to the maximum possible value of MOS ($s_i=2$ in our case) because the minimum value of MOS is zero.

\subsubsection{Coverage} 
To quantify how wide the range of the perceptual differences covered by the images of a dataset, we measure the difference between the smallest and largest perceptual difference values of the images. 
Specifically, coverage $C_i$ for gamut space $i$ is calculated as
\begin{equation}
C_i = \textrm{max}(z_i)-\textrm{min}(z_i),
\end{equation} 
where $z_i$ is a set of the normalized perceptual differences $\tilde{d_i}$ of all images in the dataset when the gamut is reduced to target gamut $i$ from the reference gamut ($i=1,\dots,N$). 
The maximum value of $C_i$ is obtained when the dataset contains images corresponding to both no-difference (MOS = 0) and clear difference (MOS = 2) for the $i$th target gamut.
In other words, one image has less or no colors outside the $i$th gamut space so that it does not cause perceptual difference by gamut reduction, but the other image contains lots of colors outside the space and thus its perceptual difference can be clearly observed by gamut reduction.

\subsubsection{Total Coverage}
This is the relative area occupied by the data points in the space of perceptual differences.
It is similar to $C_i$, but considers the interaction of different dimensions in $Z=\{z_1,z_2,\dots,z_N\}$.
It is calculated as follows:
\begin{equation}
C_{total}=\sqrt[N]{\int \textrm{convex}(Z)},
\end{equation} where $\textrm{convex}(Z)$ returns the convex hull for $N$-dimensional vectors in $Z$. 
$C_{total}$ becomes the largest when the dataset consists of images having the maximum coverage of perceptual difference for all target gamuts.
Using a dataset having a large value of $C_{total}$ in an experiment implies that images having extreme perceptual characteristics (i.e., both severe and little perceptual differences) under gamut change are employed.

\subsubsection{Uniformity}
While the above coverage measures consider the range of perceptual differences observed in the images, uniformity measures how evenly the perceptual differences are distributed within the range. 
For this, we use the information entropy, which is popularly used to measure the uniformity of a distribution. 
In other words, we construct the histogram of $z_i$, and then compute its entropy as follows in order to quantify the uniformity of the distribution of the perceptual differences.
\begin{equation}
U_i = -\sum_{k=1}^B p_{i,k}\log_B p_{i,k},
\end{equation} where $B$ is the number of bins of the histogram and $p_{i,k}$ is the ratio of the images of which perceptual differences are in the range of the $k$th bin.
The uniformity has the largest value of 1 when the perceptual differences of the dataset are uniformly distributed.
It becomes low when the dataset contains images having similar perceptual differences, and reaches 0 when the perceptual differences are the same for all images.

\subsubsection{Total Uniformity}
This measures the uniformity of perceptual differences over the whole dimensions of reduced target gamuts.
In this case, we compute the $N$-dimensional histogram of $Z$ and its entropy, i.e., 
\begin{equation}
U_{total} = -\frac{1}{N}\sum_{i=1}^N {\sum_{k=1}^B {q_{i,k}\log_B q_{i,k}}}, 
\end{equation} where $B$ is the number of bins for each dimension of the histogram and $q_{i,k}$ is the normalized count in the $k$th bin (normalized over the whole dimension).
It becomes the largest value (i.e., 1) when a dataset contains diverse images in terms of perceptual differences and the perceptual differences are uniformly distributed over all target gamuts. 
On the other hand, it has the lowest value of 0 when the dataset contains images that show the same amount of perceptual difference for all target gamuts.
A dataset having a large value of $U_{total}$ is beneficial to conduct experiments with images having diverse perceptual characteristics under gamut change.

\subsection{Analysis of Existing Datasets} \label{sec:existing}
\begin{figure*}[!t]
\centering
\subfloat[HdM]{\includegraphics[width=3in]{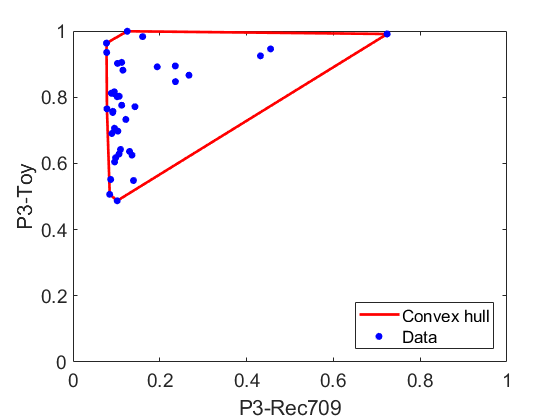}%
\label{fig_CHdM}}
\hfil
\subfloat[Arri]{\includegraphics[width=3in]{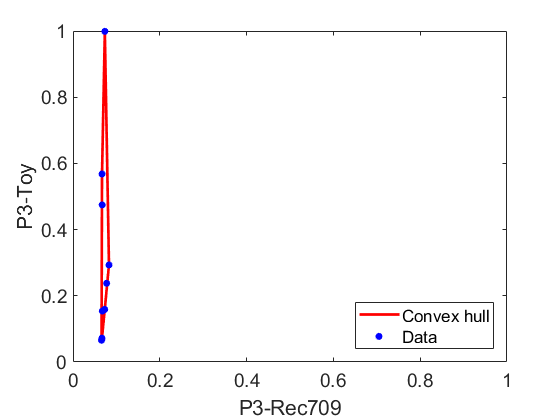}%
\label{fig_CArri}}
\caption{Measured perceptual differences and corresponding convex hulls for the (a) HdM and (b) Arri datasets.}
\label{fig_Cplot}
\end{figure*}

We analyze the two existing WCG datasets, HdM and Arri, in terms of the four criteria described above\footnote{These are the only publicly available datasets that support Rec.2020.}.
In this experiment, we collect 38 and 11 images from each dataset, respectively.
We use the perceptual difference for the 49 images due to successive gamut reduction from the reference P3 gamut to the Rec.709 and Toy gamuts as in Section~\ref{sec:fit}.
We then measure the four criteria of the two WCG datasets.
For (total) uniformity, we use 10 bins for each dimension of the histograms (i.e., $B=10$).
The measured criteria are summarized in \tablename~\ref{table_WCGDBA}.
In addition, the distributions of the perceptual difference for the two datasets are shown in \figurename~\ref{fig_Cplot}.

First, the coverages of the two datasets have different behaviors depending on the target gamuts.
The perceptual differences of the images in the HdM dataset cover over about a half of the scale for both target gamuts as shown in \figurename~\ref{fig_CHdM}.
For the case of gamut reduction to Toy, the perceptual difference is biased to large values because most images of HdM contain many pixels with highly saturated colors, which produces large perceptual difference when the gamut is reduced.
On the contrary, pixels with highly saturated color are few in the images of the Arri dataset, so the coverage criterion for Rec.709 is low while that for Toy is high as shown in \figurename~\ref{fig_CArri}.

Similarly to the results of the dimension-wise coverage criterion, the HdM dataset has a medium level of total coverage of perceptual differences, showing the convex hull covering almost the upper-half area in \figurename~\ref{fig_CHdM}.
On the other hand, although the coverage value for the Toy gamut is large as shown in \figurename~\ref{fig_CArri}, the total coverage of the Arri dataset is small due to the extremely low coverage for Rec.709.
Note that $z_{Toy}$ would be always higher than $z_{709}$ for the same image because the details of color are more distorted in Toy, so the practical maximum possible value of total coverage is 0.707 ($= \sqrt{0.5}$).

In terms of uniformity, the perceptual differences caused by the large gamut difference (i.e., the case of Toy) are quite uniformly distributed for both datasets.
For the small gamut reduction (to Rec.709), the perceptual differences of the HdM dataset are slightly biased to low values.
The perceptual differences of the Arri dataset are extremely biased, so all data points are allocated in a single bin and the uniformity is zero.

In the case of total uniformity, there exist differences between the two datasets. 
The perceptual differences of the HdM dataset are quite uniformly distributed on the two-dimensional space in \figurename~\ref{fig_CHdM}, although the data points are slightly biased to the upper region (where large perceptual differences occur due to large gamut reduction). 
For the Arri dataset, the perceptual differences are biased to the left-side in \figurename~\ref{fig_CArri}, so the total uniformity becomes low. 

Overall, each of the two datasets has its own strengths and limitations in a complementary manner.
HdM has a relatively small coverage of $z_{Toy}$, while Arri has limited characteristics in the Rec.709 gamut.
For example, if the Arri dataset is used for an experiment involving gamut changes, the experiment would draw biased conclusion for small gamut difference.
Based on this understanding, one can choose either of the two datasets for particular research problems; for instance, the Arri dataset could be more effective for the experiments that focus on large gamut difference.
Furthermore, one can obtain an enhanced dataset by supplementing one of the two datasets with particular contents having characteristics desired for the given objective.

\begin{table}[!t]
\renewcommand{\arraystretch}{1.3}
\caption{Results of WCG Dataset Characterization}
\label{table_WCGDBA}
\centering
\scalebox{0.95}{
\begin{tabular}{lccccccc}
\toprule
\multirow{2}{*}{\textbf{Criteria}} & \multicolumn{3}{c}{\textbf{HdM}} & \phantom{} & \multicolumn{3}{c}{\textbf{Arri}} \\
\cmidrule{2-4} \cmidrule{6-8} 
&\textbf{Toy}&\textbf{Rec.709}&\textbf{Total}&&\textbf{Toy}&\textbf{Rec.709}&\textbf{Total}\\
\midrule
Coverage & 0.512 & 0.647 & 0.412 && 0.933 & 0.017 & 0.093\\
Uniformity & 0.707 & 0.509 & 0.550 && 0.713 & 0.000 & 0.357\\
\bottomrule
\end{tabular}
}
\end{table}

\section{Application to evaluation of gamut mapping algorithms} \label{sec:scenario}
In this section, we present another practical application of the proposed framework, which is the problem of evaluation of GMAs.
In this scenario, the proposed framework plays a role to select image contents used for performance comparison of different GMAs. We demonstrate the reliability of the framework for selection of representative contents for fair comparison.

\subsection{Scenario}
The main goal of the scenario is to benchmark performance of GMAs.
Each GMA is applied to a set of source image contents having wide gamuts, and its performance is measured by an objective quality metric in terms of perceptual color information loss in the gamut-reduced images in comparison to the original ones. 
Here, which image dataset is used is an important issue. 
For instance, if images that do not have color profiles challenging enough to reveal distinguished gamut mapping performance, the GMAs may be evaluated to perform similarly, which may not be the case if challenging images are included. 
Therefore, careful selection of the images is required to obtain unbiased benchmarking results, for which the proposed framework can be used. 
Therefore, our objective is to evaluate the reliability of the benchmarking results between different source content selection methods. 

We limit the number of GMAs for comparison to two in order to validate the effectiveness of the proposed framework clearly rather than to present extensive benchmarking of many GMAs.
One is the state-of-the-art gamut reduction algorithm~\cite{GRA} that adaptively modifies local contrast of pixels residing outside of the target gamut based on the Retinex theory~\cite{Retinex}. 
For the other one, we use the gamut compression algorithm~\cite{CIEguidelines} that maps the entire color of the source image inside the target gamut in the CIE 1931 space.

To evaluate the performance of gamut mapping, we use the color image difference (CID)~\cite{CID} that predicts perceptual color difference between the reference and gamut-reduced image, which is used to evaluate performance of the gamut reduction algorithm in~\cite{GRA}.
As the main objective of conducting the scenario, we focus on the reliability and robustness of test results with representative contents selected by our framework.
First, the selected dataset should sufficiently cover diverse gamut characteristics so that it is representative.
Second, in terms of robustness, experiments with content selection followed by the same procedure should produce consistent results and conclusions regardless of repetition.

\subsection{Content Selection}
The pool of candidate source images consists of half-HD ($960\times1080$ pixels) WCG images from both the HdM and Arri datasets. 
After excluding images containing no or too few pixels in WCG (outside the Rec.709 gamut) from the data used in Section~\ref{sec:existing}, 35 candidate images are used.
The reference gamut is Rec.2020, and we use three target gamuts for gamut mapping: P3, Rec.709, and Toy.

The proposed framework is applied to select representative images from the pool. 
As described in Section~\ref{sec:fit}, each candidate image is represented by a two-dimensional perceptual feature vector. 
Then, the $k$-means clustering algorithm with $k=3$ is used to group them into three clusters, from each of which three images are randomly selected. 
For comparison, content selection using an existing content feature, colorfulness~\cite{Colorfulness}, is also conducted. 
It measures the variety and intensity of colors in an image. 
The colorfulness features computed for the candidate images are also clustered into three groups and three images are randomly chosen from each group. 
These content selection procedures are repeated 100 times with different random seeds.

\subsection{Evaluation}
In order to compare the two GMAs, we define CID gain $g_t$ for target gamut $t$ and for a source image as
\begin{equation}
g_t=\textrm{CID}(I_0,\textrm{GC}(I_0,t)) - \textrm{CID}(I_0,\textrm{GR}(I_0,t)),
\end{equation} where $I_0$ is the reference image, and $\textrm{GC}(I_0,t)$ and $\textrm{GR}(I_0,t)$ are the gamut-compressed and gamut-reduced versions of $I_0$, respectively.
$g_t$ becomes positive when the gamut reduction algorithm performs better than the gamut compression algorithm, and its absolute value indicates the degree of the performance difference.

Using the CID gains for 100 repetitions, the two content selection methods are compared with respect to two aspects: robustness and representativeness. 
First, a content selection method is considered to be robust when the CID gains remain consistent, i.e., the averages and standard deviations of the CID gains over the selected images are similar across the repetitions. 
Second, a dataset of images chosen by a content selection method is regarded as being representative if the images have diverse color characteristics. 
Thus, the CID gains lie in a wide range, resulting in a large average and standard deviation over the images.

\subsection{Results}
\begin{figure*}[!t]
\centering
\subfloat[Framework (P3)]{\includegraphics[width=2.3in]{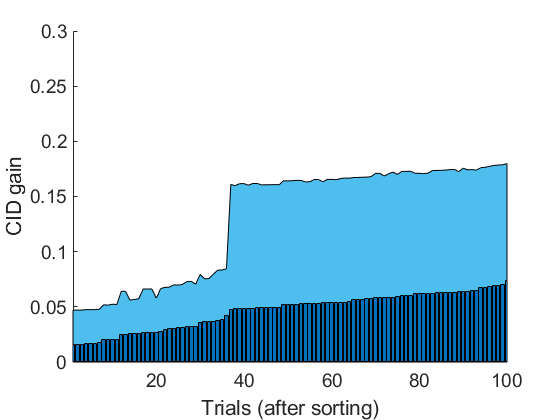}%
\label{fig_CID_F_P3}}
\hfil
\subfloat[Framework (Rec.709)]{\includegraphics[width=2.3in]{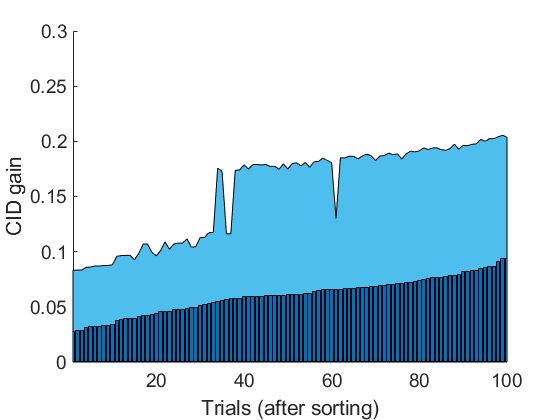}%
\label{fig_CID_F_709}}
\hfil
\subfloat[Framework (Toy)]{\includegraphics[width=2.3in]{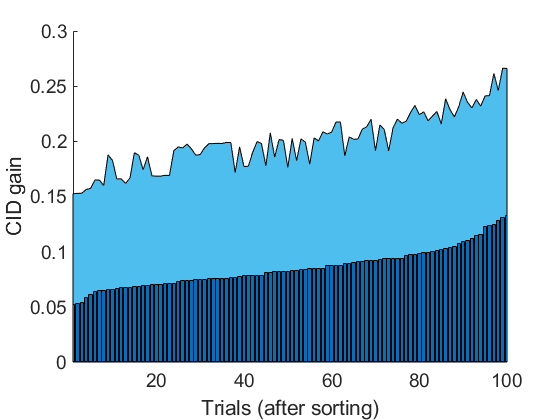}%
\label{fig_CID_F_Toy}}
\hfil
\subfloat[Colorfulness (P3)]{\includegraphics[width=2.3in]{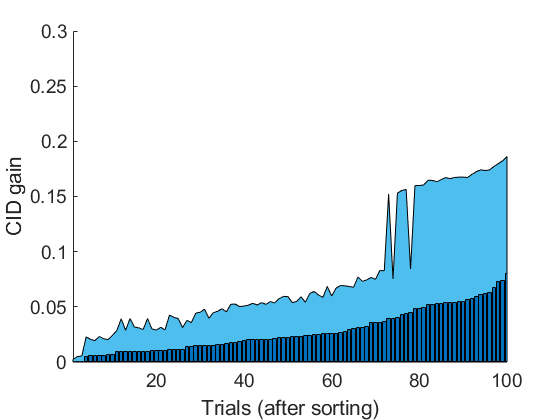}%
\label{fig_CID_C_P3}}
\hfil
\subfloat[Colorfulness (Rec.709)]{\includegraphics[width=2.3in]{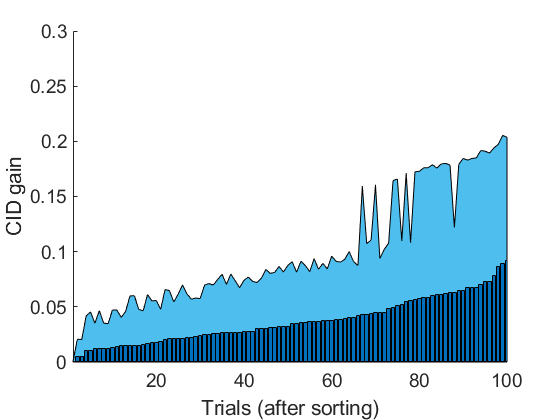}%
\label{fig_CID_C_709}}
\hfil
\subfloat[Colorfulness (Toy)]{\includegraphics[width=2.3in]{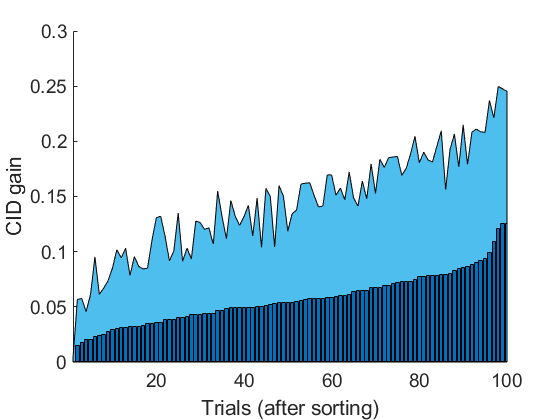}%
\label{fig_CID_C_Toy}}
\caption{CID gains for the images selected by the method using the proposed framework or the one using colorfulness. The average and standard deviation values of the CID gains over the selected images are represented by the bars with dark color and the shaded area, respectively.}
\label{fig_CID_gain}
\end{figure*}

\figurename~\ref{fig_CID_gain} shows the average and standard deviation of CID gains for the selected images with respect to the target gamut and selection method.
In all cases, the average CID gains are positive, which indicates that the gamut reduction algorithm produces gamut-reduced images with smaller difference from the reference ones compared to the gamut compression algorithm.
When the three target gamuts are compared, a smaller gamut yields larger CID gains because more color distortion is introduced by the gamut compression algorithm than the gamut reduction algorithm as the gamut difference becomes larger.

The two selection methods show clearly distinct results. 
First, the average and standard deviation of the CID gains appear more similar across 100 trials when the proposed framework is used, particularly when the target gamut is small. 
In order to statistically assess this, we conduct one-sided F-tests under the null hypothesis that the two populations (one for the proposed framework and the other for the method using colorfulness) of the average (or standard deviation) values of the CID gains have the same variance. 
The results are shown in \tablename~\ref{table_test}, which confirms that the cases involving large gamut changes show statistically significant difference (i.e., Rec.709 and Toy for the average and Toy for the standard deviation).
Note that for P3, the gamut difference from Rec.2020 is small, so the average and standard deviation of the CID gains are also small. 
These results demonstrate that the selection method has an impact on the results of GMA comparison, where content selection using the proposed framework provides improved robustness.

Second, on average, the average and standard deviation values are larger for the case using the proposed framework than for the case using colorfulness. 
Since many images in the pool are not challenging for GMAs as shown in Section~\ref{sec:existing}, for which the CID gain is small, a larger average or standard deviation value indicates a more representative dataset. 
We perform one-sided t-tests under the null hypothesis that the two populations of the average (or standard deviation) values of the CID gains have the same mean. 
As shown in \tablename~\ref{table_test}, the null hypothesis is rejected in all cases, indicating that the average and standard deviation values are significantly larger for our method. 
This confirms representativeness of the dataset obtained using our method and, consequently, reliability of the results of the benchmarking.

\begin{table}[!t]
\renewcommand{\arraystretch}{1.3}
\caption{Result of the Statistical Tests Comparing the Selection Method Using the Proposed Framework and the One Using Colorfulness. The Degrees of Freedom of t-tests on Average and Standard Deviation Using the Welch-Satterthwaite Equation Are 181.8 and 134.8, Respectively. Statistical Significance (Bonferroni-corrected for Multiple Comparison) Is Marked in Bold.}
\label{table_test}
\centering
\scalebox{0.95}{
\begin{tabular}{lccccc}
\toprule
\multirow{2}{*}{Target gamut} & \multicolumn{2}{c}{Average} & \phantom{} & \multicolumn{2}{c}{Standard deviation} \\ 
\cmidrule{2-3} \cmidrule{5-6}
& Statistics& $p$-value && Statistics& $p$-value\\ \midrule
\textbf{F-tests} &&&&&\\
\hspace{0.3cm}P3& $F=0.72$& $0.050$&&$F=0.94$& $0.385$\\
\hspace{0.3cm}Rec.709& $F=0.63$& \boldmath$0.012$&& $F=0.72$& $0.053$\\
\hspace{0.3cm}Toy& $F=0.54$& \boldmath$0.001$&& $F=0.19$& \boldmath$<0.001$\\ %
\textbf{t-tests} &&&&\\
\hspace{0.3cm}P3& $t=7.40$& \boldmath$<0.001$&&$t=6.43$& \boldmath$<0.001$\\
\hspace{0.3cm}Rec.709& $t=8.79$& \boldmath$<0.001$&& $t=7.50$& \boldmath$<0.001$\\
\hspace{0.3cm}Toy& $t=9.65$& \boldmath$<0.001$&& $t=9.43$& \boldmath$<0.001$\\ \bottomrule
\end{tabular}
}
\end{table}

For comparison, we provide further results using selection features other than colorfulness. 
We use two no-reference color quality metrics: contrast enhancement based contrast-changed image quality measure (CEIQ)~\cite{CEIQ} and accelerated screen image quality evaluator (ASIQE)~\cite{ASIQE}. 
The former is a metric based on a learned support vector machine using multiple features estimating contrast distortion, while the latter assesses image quality considering four types of quality features consisting of picture complexity, screen content statistics, global brightness quality, and sharpness of details. 
We conduct statistical tests comparing the CID gains obtained by our framework and the method using either CEIQ or ASIQE. 
The results are shown in \tablename~\ref{table_test_new}. 
Similar to the results in \tablename~\ref{table_test} using colorfulness, statistical significance is observed for F-tests in the cases of the large gamut change (i.e. between Rec.709 and Toy) and for t-tests in all cases. Thus, our framework can effectively select representative contents reliably compared to the methods using these image quality metrics.

\begin{table}[!t]
\renewcommand{\arraystretch}{1.3}
\caption{Result of the Statistical Tests Comparing the Selection Method Using the Proposed Framework and the Ones Using CEIQ and ASIQE. Statistical Significance Is Marked in Bold.}
\label{table_test_new}
\centering
\scalebox{0.95}{
\begin{tabular}{lccccc}
\toprule
\multirow{2}{*}{Target gamut} & \multicolumn{2}{c}{Average} & \phantom{} & \multicolumn{2}{c}{Standard deviation} \\ 
\cmidrule{2-3} \cmidrule{5-6}
& Statistics& $p$-value && Statistics& $p$-value\\ \midrule
\multicolumn{2}{l}{\textbf{F-tests (CEIQ)}} &&&&\\
\hspace{0.3cm}P3& $F=0.78$& $0.103$&&$F=0.83$& $0.170$\\
\hspace{0.3cm}Rec.709& $F=0.66$& \boldmath$0.020$&& $F=0.57$& \boldmath$<0.001$\\
\hspace{0.3cm}Toy& $F=0.49$& \boldmath$<0.001$&& $F=0.14$& \boldmath$<0.001$\\ %
\textbf{t-tests} &&&&\\
\hspace{0.3cm}P3& $t=10.73$& \boldmath$<0.001$&&$t=10.44$& \boldmath$<0.001$\\
\hspace{0.3cm}Rec.709& $t=12.04$& \boldmath$<0.001$&& $t=11.80$& \boldmath$<0.001$\\
\hspace{0.3cm}Toy& $t=12.23$& \boldmath$<0.001$&& $t=12.78$& \boldmath$<0.001$\\ \midrule
\multicolumn{2}{l}{\textbf{F-tests (ASIQE)}}&&&&\\
\hspace{0.3cm}P3& $F=0.76$& $0.082$&&$F=0.92$& $0.340$\\
\hspace{0.3cm}Rec.709& $F=0.57$& \boldmath$0.002$&& $F=0.62$& \boldmath$0.009$\\
\hspace{0.3cm}Toy& $F=0.44$& \boldmath$<0.001$&& $F=0.11$& \boldmath$<0.001$\\ %
\textbf{t-tests} &&&&\\
\hspace{0.3cm}P3& $t=9.77$& \boldmath$<0.001$&&$t=9.27$& \boldmath$<0.001$\\
\hspace{0.3cm}Rec.709& $t=10.81$& \boldmath$<0.001$&& $t=11.00$& \boldmath$<0.001$\\
\hspace{0.3cm}Toy& $t=10.90$& \boldmath$<0.001$&& $t=13.24$& \boldmath$<0.001$\\ \bottomrule
\end{tabular}
}
\end{table}

\section{Conclusion} \label{sec:conclusion}
We proposed a content characterization method for a WCG image content and evaluated it in practical applications.
The main idea was to obtain perceptual color differences due to successive gamut reduction as content characteristics for the WCG content.
As one of the practical use cases of the framework, we analyzed existing datasets by measuring dataset characterization criteria on the WCG characteristics.
Four criteria consisting of coverage, total coverage, uniformity, and total uniformity effectively characterized WCG datasets.
In addition, we validated WCG content characteristics as a content selection feature in a GMA benchmarking scenario.
Using the framework, we were able to select representative WCG contents, and draw robust and reliable benchmarking results.

In the future, the proposed framework can be improved in several ways. 
First, we employed $cssim$ for objective quality assessment due to its superiority. 
If metrics that perform better than $cssim$ are developed in the future, e.g., deep learning-based methods, our framework could benefit from employing such improved metrics. 
Second, the scope of the framework could be extended to video contents by considering the temporal dimension of color perception.

\ifCLASSOPTIONcaptionsoff
  \newpage
\fi

\bibliographystyle{bibtex/bib/IEEEtran}
\nocite{}
\bibliography{refs}

\begin{thebibliography}{10}
\providecommand{\url}[1]{#1}
\csname url@samestyle\endcsname
\providecommand{\newblock}{\relax}
\providecommand{\bibinfo}[2]{#2}
\providecommand{\BIBentrySTDinterwordspacing}{\spaceskip=0pt\relax}
\providecommand{\BIBentryALTinterwordstretchfactor}{4}
\providecommand{\BIBentryALTinterwordspacing}{\spaceskip=\fontdimen2\font plus
\BIBentryALTinterwordstretchfactor\fontdimen3\font minus
  \fontdimen4\font\relax}
\providecommand{\BIBforeignlanguage}[2]{{%
\expandafter\ifx\csname l@#1\endcsname\relax
\typeout{** WARNING: IEEEtran.bst: No hyphenation pattern has been}%
\typeout{** loaded for the language `#1'. Using the pattern for}%
\typeout{** the default language instead.}%
\else
\language=\csname l@#1\endcsname
\fi
#2}}
\providecommand{\BIBdecl}{\relax}
\BIBdecl

\bibitem{ICIP2018}
J.~Lee, T.~Vigier, P.~{Le Callet}, and J.-S. Lee, ``A perception-based
  framework for wide color gamut content selection,'' in \emph{Proceedings of
  IEEE International Conference on Image Processing}, Oct. 2018, pp. 709--713.

\bibitem{ITU709}
ITU, ``{ITU-R BT.709-6. Parameter values for the HDTV standards for production
  and international programme exchange},'' Tech. Rep., 2015.

\bibitem{ITU2012}
------, ``{ITU-R BT.2020-2. Parameter values for ultra-high definition
  television systems for production and international programme exchange},''
  Tech. Rep., 2015.

\bibitem{CIE1931}
T.~Smith and J.~Guild, ``The {C.I.E.} colorimetric standards and their use,''
  \emph{Transactions of the Optical Society}, vol.~33, no.~3, pp. 73--134,
  1931.

\bibitem{UHDTV}
C.~Chinnock, ``The status of wide color gamut {UHD-TV}s,'' Tech. Rep., 2016.

\bibitem{CIEguidelines}
{CIE}, ``Guidelines for the evaluation of gamut mapping algorithms,'' Tech.
  Rep., 2004.

\bibitem{GRA}
S.~W. Zamir, J.~Vazquez-Corral, and M.~Bertalmio, ``Gamut mapping in
  cinematography through perceptually-based contrast modification,'' \emph{IEEE
  Journal of Selected Topics in Signal Processing}, vol.~8, no.~3, pp.
  490--503, 2014.

\bibitem{Retinex}
E.~H. Land and J.~J. McCann, ``Lightness and retinex theory,'' \emph{Journal of
  the Optical Society of America}, vol.~61, no.~1, pp. 1--11, 1971.

\bibitem{Saliency}
K.~{Gu}, S.~{Wang}, H.~{Yang}, W.~{Lin}, G.~{Zhai}, X.~{Yang}, and W.~{Zhang},
  ``Saliency-guided quality assessment of screen content images,'' \emph{IEEE
  Transactions on Multimedia}, vol.~18, no.~6, pp. 1098--1110, Jun. 2016.

\bibitem{Blind2016}
K.~{Gu}, S.~{Wang}, G.~{Zhai}, S.~{Ma}, X.~{Yang}, W.~{Lin}, W.~{Zhang}, and
  W.~{Gao}, ``Blind quality assessment of tone-mapped images via analysis of
  information, naturalness, and structure,'' \emph{IEEE Transactions on
  Multimedia}, vol.~18, no.~3, pp. 432--443, Mar. 2016.

\bibitem{Sampling}
Z.~{Fan}, T.~{Jiang}, and T.~{Huang}, ``Active sampling exploiting reliable
  informativeness for subjective image quality assessment based on pairwise
  comparison,'' \emph{IEEE Transactions on Multimedia}, vol.~19, no.~12, pp.
  2720--2735, Dec. 2017.

\bibitem{Blind2018}
X.~{Min}, K.~{Gu}, G.~{Zhai}, J.~{Liu}, X.~{Yang}, and C.~W. {Chen}, ``Blind
  quality assessment based on pseudo-reference image,'' \emph{IEEE Transactions
  on Multimedia}, vol.~20, no.~8, pp. 2049--2062, Aug. 2018.

\bibitem{IQA2018}
P.~G. {Freitas}, W.~Y.~L. {Akamine}, and M.~C.~Q. {Farias}, ``No-reference
  image quality assessment using orthogonal color planes patterns,'' \emph{IEEE
  Transactions on Multimedia}, vol.~20, no.~12, pp. 3353--3360, Dec. 2018.

\bibitem{GMA1988}
M.~C. Stone, W.~B. Cowan, and J.~C. Beatty, ``Color gamut mapping and the
  printing of digital color images,'' \emph{ACM Transactions on Graphics},
  vol.~7, no.~4, pp. 249--292, Oct. 1988.

\bibitem{GMA1989}
G.~M. Murch and J.~M. Taylor, ``Color in computer graphics: Manipulating and
  matching color,'' in \emph{Advances in Computer Graphics V}.\hskip 1em plus
  0.5em minus 0.4em\relax Springer, 1989, pp. 19--47.

\bibitem{GMA1997}
F.~Ebner and M.~D. Fairchild, ``Gamut mapping from below: Finding minimum
  perceptual distances for colors outside the gamut volume,'' \emph{Color
  Research \& Application}, vol.~22, no.~6, pp. 402--413, 1997.

\bibitem{GMA1997M}
J.~Morovic and M.~R. Luo, ``Gamut mapping algorithms based on psychophysical
  experiment,'' in \emph{Proceedings of the Color and Imaging Conference}, vol.
  1997, no.~1, 1997, pp. 44--49.

\bibitem{GMA1999}
S.~O. Naoya~Katoh, Masahiko~Ito, ``Three-dimensional gamut mapping using
  various color difference formulae and color spaces,'' \emph{Journal of
  Electronic Imaging}, vol.~8, no.~4, pp. 365--379, 1999.

\bibitem{spatial2003}
J.~Morovic and Y.~Wang, ``A multi--resolution, full--colour spatial gamut
  mapping algorithm,'' in \emph{Proceedings of the Color and Imaging
  Conference}, vol. 2003, no.~1, 2003, pp. 282--287.

\bibitem{spatial2006}
P.~Zolliker and K.~Simon, ``Adding local contrast to global gamut mapping
  algorithms,'' in \emph{Proceedings of the Conference on Colour in Graphics,
  Imaging, and Vision}, vol. 2006, no.~1, 2006, pp. 257--261.

\bibitem{spatial2007}
I.~Farup, C.~Gatta, and A.~Rizzi, ``A multiscale framework for spatial gamut
  mapping,'' \emph{IEEE Transactions on Image Processing}, vol.~16, no.~10, pp.
  2423--2435, 2007.

\bibitem{Retaining2007}
P.~{Zolliker} and K.~{Simon}, ``Retaining local image information in gamut
  mapping algorithms,'' \emph{IEEE Transactions on Image Processing}, vol.~16,
  no.~3, pp. 664--672, Mar. 2007.

\bibitem{Efficient2007}
{\O}.~Kol{\aa}s and I.~Farup, ``Efficient hue-preserving and edge-preserving
  spatial color gamut mapping,'' in \emph{Proceedings of the Color and Imaging
  Conference}, 2007, pp. 207--212.

\bibitem{Spatial2012}
A.~Alsam and I.~Farup, ``Spatial colour gamut mapping by orthogonal projection
  of gradients onto constant hue lines,'' in \emph{Advances in Visual
  Computing}, 2012, pp. 556--565.

\bibitem{Gamut2017}
C.~Gatta and I.~Farup, ``Gamut mapping in {RGB} colour spaces with the
  iterative ratios diffusion algorithm,'' in \emph{Proceedings of the IS\&T
  International Symposium on Electronic Imaging}, 2017, pp. 12--20.

\bibitem{Perceptual2008}
F.~Dugay, I.~Farup, and J.~Y. Hardeberg, ``Perceptual evaluation of color gamut
  mapping algorithms,'' \emph{Color Research \& Application}, vol.~33, no.~6,
  pp. 470--476, 2008.

\bibitem{CIELAB}
CIE, ``Recommendations on uniform color spaces, color-difference equations,
  psychometric color terms,'' Tech. Rep., 1978.

\bibitem{SCIELAB}
X.~Zhang and B.~A. Wandell, ``A spatial extension of {CIELAB} for digital color
  image reproduction,'' in \emph{SID International Symposium Digest of
  Technical Papers}, vol.~27, 1996, pp. 731--734.

\bibitem{iCAM}
M.~D.~Fairchild and G.~M.~Johnson, ``The {iCAM} framework for image appearance,
  image differences, and image quality,'' \emph{Journal of Electronic Imaging},
  vol.~13, pp. 126--138, Jan. 2004.

\bibitem{hueangle}
G.~Hong and M.~R. Luo, ``Perceptually-based color difference for complex
  images,'' in \emph{Proceedings of the Congress of the International Colour
  Association}, vol. 4421, 2002, pp. 618--622.

\bibitem{UIQ}
Z.~Wang and A.~C. {Bovik}, ``A universal image quality index,'' \emph{IEEE
  Signal Processing Letters}, vol.~9, no.~3, pp. 81--84, Mar. 2002.

\bibitem{Evaluation2006}
N.~Bonnier, F.~Schmitt, H.~Brettel, and S.~Berche, ``Evaluation of spatial
  gamut mapping algorithms,'' in \emph{Proceedings of Color and Imaging
  Conference}, 2006, pp. 56--61.

\bibitem{Evaluating2008}
J.~Y. Hardeberg, E.~Bando, and M.~Pedersen, ``Evaluating colour image
  difference metrics for gamut-mapped images,'' \emph{Coloration Technology},
  vol. 124, no.~4, pp. 243--253, 2008.

\bibitem{SHAME}
M.~Pedersen and J.~Y. Hardeberg, ``A new spatial hue angle metric for
  perceptual image difference,'' in \emph{In Proceedings of the Computational
  Color Imaging Workshop}, 2009, pp. 81--90.

\bibitem{CID}
I.~Lissner, J.~Preiss, P.~Urban, M.~S. Lichtenauer, and P.~Zolliker,
  ``Image-difference prediction: From grayscale to color,'' \emph{IEEE
  Transactions on Image Processing}, vol.~22, no.~2, pp. 435--446, 2013.

\bibitem{Physiological2016}
D.~Darcy, E.~Gitterman, A.~Brandmeyer, S.~Daly, and P.~Crum, ``Physiological
  capture of augmented viewing states: objective measures of high-dynamic-range
  and wide-color-gamut viewing experiences,'' in \emph{Proceedings of the IS\&T
  International Symposium on Human Vision and Electronic Imaging}, 2016, pp.
  HVEI126:1--9.

\bibitem{Winkler}
S.~Winkler, ``Analysis of public image and video databases for quality
  assessment,'' \emph{IEEE Journal of Selected Topics in Signal Processing},
  vol.~6, no.~6, pp. 616--625, 2012.

\bibitem{BVI}
F.~{Zhang}, F.~M. {Moss}, R.~{Baddeley}, and D.~R. {Bull}, ``{BVI-HD}: A video
  quality database for {HEVC} compressed and texture synthesized content,''
  \emph{IEEE Transactions on Multimedia}, vol.~20, no.~10, pp. 2620--2630, Oct.
  2018.

\bibitem{MM2019}
A.~{Mackin}, F.~{Zhang}, and D.~R. {Bull}, ``A study of high frame rate video
  formats,'' \emph{IEEE Transactions on Multimedia}, vol.~21, no.~6, pp.
  1499--1512, Jun. 2019.

\bibitem{Selecting2013}
M.~H. Pinson, M.~Barkowsky, and P.~{Le Callet}, ``Selecting scenes for {2D} and
  {3D} subjective video quality tests,'' \emph{EURASIP Journal on Image and
  Video Processing}, vol. 2013, no.~1, pp. 50:1--12, Aug. 2013.

\bibitem{Krasula_Features}
L.~Krasula, K.~Fliegel, P.~{Le Callet}, and M.~Kl{\'\i}ma, ``Objective
  evaluation of naturalness, contrast, and colorfulness of tone-mapped
  images,'' in \emph{Proceedings of the Applications of Digital Image
  Processing XXXVII}, vol. 9217, 2014, pp. 92\,172D:1--10.

\bibitem{LF2017}
P.~{Paudyal}, J.~{Gutiérrez}, P.~{Le Callet}, M.~{Carli}, and F.~{Battisti},
  ``Characterization and selection of light field content for perceptual
  assessment,'' in \emph{Proceedings of the International Conference on Quality
  of Multimedia Experience}, 2017, pp. 1--6.

\bibitem{Narwaria_QoMEX2014}
M.~Narwaria, C.~Mantel, M.~{Perreira Da Silva}, P.~{Le Callet}, and
  S.~Forchhammer, ``An objective method for high dynamic range source content
  selection,'' in \emph{Proceedings of the International Workshop on Quality of
  Multimedia Experience}, 2014, pp. 13--18.

\bibitem{Krasula_JSTSP}
L.~Krasula, M.~Narwaria, K.~Fliegel, and P.~{Le Callet}, ``Preference of
  experience in image tone-mapping: dataset and framework for objective
  measures comparison,'' \emph{IEEE Journal of Selected Topics in Signal
  Processing}, vol.~11, no.~1, pp. 64--74, 2017.

\bibitem{HdM}
J.~Froehlich, S.~Grandinetti, B.~Eberhardt, S.~Walter, A.~Schilling, and
  H.~Brendel, ``Creating cinematic wide gamut {HDR}-video for the evaluation of
  tone mapping operators and {HDR}-displays,'' in \emph{Proceedings of the
  Digital Photography X}, vol. 9023, 2014, pp. 90\,230X:1--10.

\bibitem{gamutex2017}
S.~W. Zamir, J.~Vazquez-Corral, and M.~Bertalm{\'\i}o, ``Gamut extension for
  cinema,'' \emph{IEEE Transactions on Image Processing}, vol.~26, no.~4, pp.
  1595--1606, 2017.

\bibitem{paired2014}
J.-S. Lee, ``On designing paired comparison experiments for subjective
  multimedia quality assessment,'' \emph{IEEE Transactions on Multimedia},
  vol.~16, no.~2, pp. 564--571, 2014.

\bibitem{ITU500}
ITU, ``{ITU-R BT.500-13. Methodology for the subjective assessment of the
  quality of television pictures},'' Tech. Rep., 2012.

\bibitem{CD2016}
B.~Ortiz-Jaramillo, A.~Kumcu, and W.~Philips, ``Evaluating color difference
  measures in images,'' in \emph{Proceedings of the International Conference on
  Quality of Multimedia Experience}, 2016, pp. 1--6.

\bibitem{cssim2003}
A.~Toet and M.~P. Lucassen, ``A new universal colour image fidelity metric,''
  \emph{Displays}, vol.~24, no.~4, pp. 197--207, 2003.

\bibitem{CIEDE2000}
G.~Sharma, W.~Wu, and E.~Daa, ``The {CIEDE2000} color-dierence formula:
  Implementation notes, supplementary test data, and mathematical
  observations,'' \emph{Color Research \& Application}, vol.~30, pp. 21--30,
  2004.

\bibitem{CIEDES}
X.~Zhang and B.~A. Wandell, ``A spatial extension of {CIELAB} for digital
  color-image reproduction,'' \emph{Journal of the Society for Information
  Display}, vol.~5, no.~1, pp. 61--63, 1997.

\bibitem{Colorfulness}
D.~Hasler and S.~S{\"u}sstrunk, ``Measuring colourfulness in natural images,''
  in \emph{Proceedings of the IST/SPIE Electronic Imaging 2003: Human Vision
  and Electronic Imaging VIII}, vol. 5007, no.~19, 2003, pp. 87--95.

\bibitem{ChromaEx}
M.~H. Pinson and S.~Wolf, ``A new standardized method for objectively measuring
  video quality,'' \emph{IEEE Transactions on Broadcasting}, vol.~50, no.~3,
  pp. 312--322, 2004.

\bibitem{ColorAppear}
G.~M. Johnson, ``Using color appearance in image quality metrics,'' in
  \emph{Proceedings of the Second International Workshop on Video Processing
  and Quality Metrics for Consumer Electronics}, 2006.

\bibitem{JCND}
C.~H. Chou and K.~C. Liu, ``A fidelity metric for assessing visual quality of
  color images,'' in \emph{Proceedings of the 16th International Conference on
  Computer Communications and Networks}, 2007, pp. 1154--1159.

\bibitem{AdaptiveSpatio}
U.~Rajashekar, Z.~Wang, and E.~P. Simoncelli, ``Quantifying color image
  distortions based on adaptive spatio-chromatic signal decompositions,'' in
  \emph{Proceedings of the 16th IEEE International Conference on Image
  Processing}, 2009, pp. 2213--2216.

\bibitem{ITU149}
ITU, ``{ITU-T J.149. Method for specifying accuracy and cross-calibration of
  Video Quality Metrics (VQM)},'' Tech. Rep., 2004.

\bibitem{CEIQ}
J.~Yan, J.~Li, and X.~Fu, ``No-reference quality assessment of
  contrast-distorted images using contrast enhancement,'' \emph{arXiv preprint
  arXiv:1904.08879}, pp. 1--15, 2019.

\bibitem{ASIQE}
K.~{Gu}, J.~{Zhou}, J.~{Qiao}, G.~{Zhai}, W.~{Lin}, and A.~C. {Bovik},
  ``No-reference quality assessment of screen content pictures,'' \emph{IEEE
  Transactions on Image Processing}, vol.~26, no.~8, pp. 4005--4018, 2017.

\end{thebibliography}

\begin{IEEEbiography}[{\includegraphics[width=1in,height=1.25in,clip,keepaspectratio]{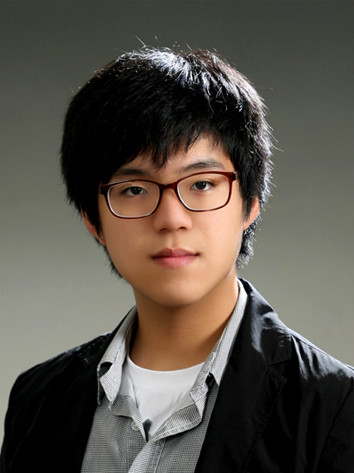}}]{Junghyuk Lee}
received his B.S. degree from the School of Integrated Technology at Yonsei Uni- versity, Korea, in 2015, where he is currently working toward the Ph.D. degree. His research interests include multimedia signal processing and wide color gamut imaging.
\end{IEEEbiography}

\begin{IEEEbiography}[{\includegraphics[width=1in,height=1.25in,clip,keepaspectratio]{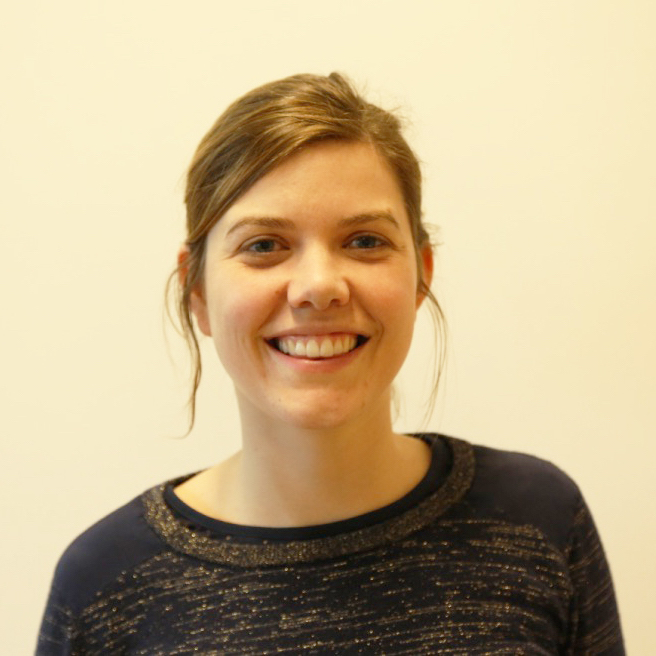}}]{Toinon Vigier}
obtained a PhD in July 2015 from the Ecole Centrale de Nantes in the Ambiances Architectures and Urbanity lab, where she focused on virtual reality for urban studies. She specifically studied the impact of rendering and color effects on the perception of urban atmospheres through VR subjective tests. She was then a postdoctoral fellow in the Image Video and Communication team at Universit\'e de Nantes. She worked mainly on video quality and eye-tracking studies in the European CATRENE project UltraHD-4U which aims at studying and implementing a complete chain for the broadcasting of UHD-4K videos. Since September 2016, she is an Associate Professor at Université de Nantes in the Image Perception Interaction research team of the Laboratory of Digital Sciences in Nantes (LS2N). Her research mainly focuses on the study, the analysis and the prediction of the quality of experience for immersive and interactive multimedia through subjective and objective measures. She is currently involved in various national and international interdisciplinary projects focusing on user experience in immersive VR media for various applications (health, cinema, architecture, design…). She is also active in the standardization working group IEEE 3333.1 and she served as reviewers in a lot of international conferences and journals (IEEE TIP, IEEE TCSVT, SPIE JEI, IEEE VR, IEEE QoMEX, ACM TVX, IEEE MMSP). 
\end{IEEEbiography}

\begin{IEEEbiography}[{\includegraphics[width=1in,height=1.25in,clip,keepaspectratio]{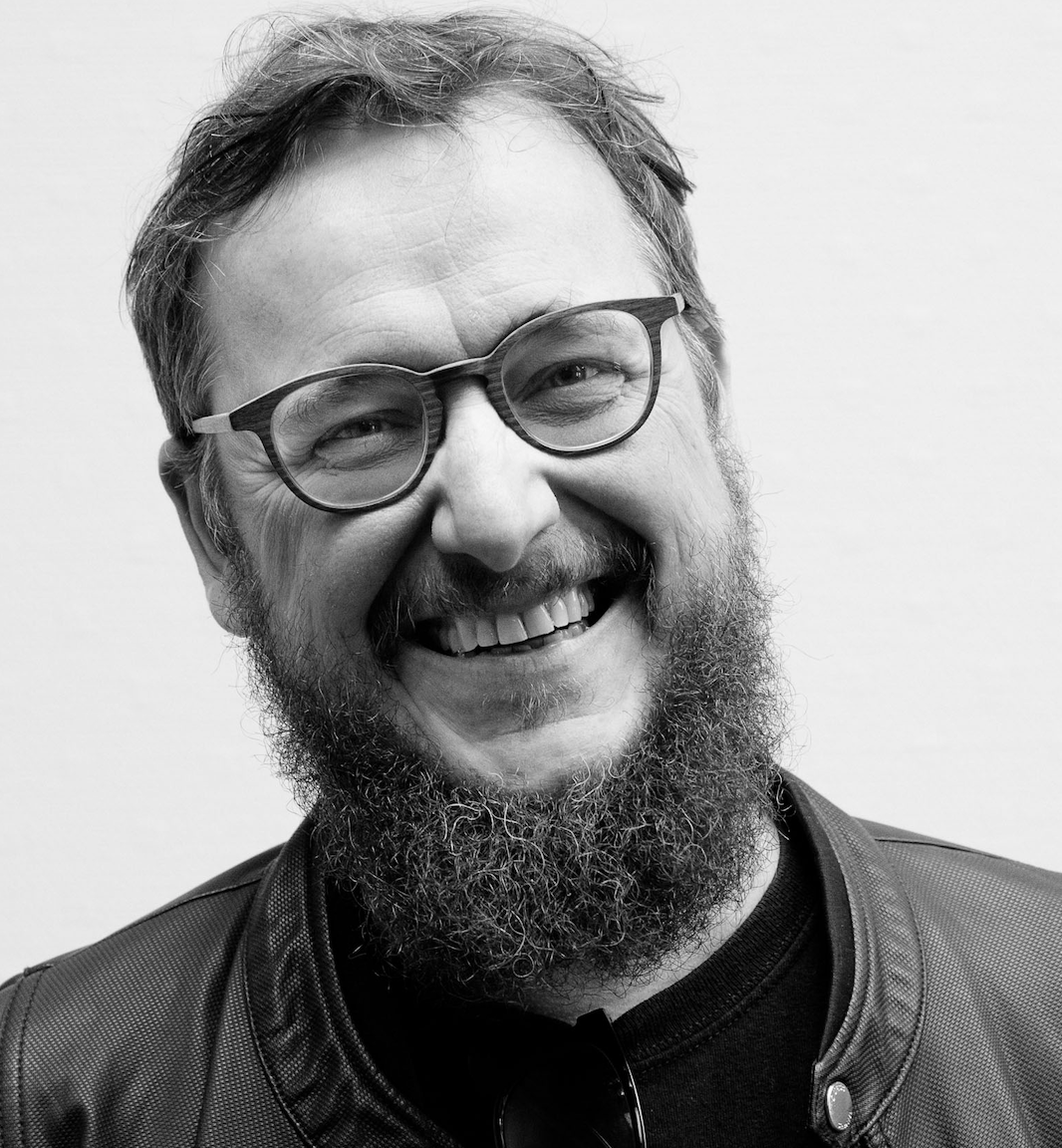}}]{Patrick Le Callet}
(IEEE Fellow) is full professor at University of Nantes, in the Electrical Engineering and the Computer Science departments of Polytech Nantes. He is one of the steering director of CNRS LS2N lab (450 researchers). He is also the scientific director of the cluster “Ouest Industries Cr\'eatives”, gathering more than 10 institutions (including 3 universities). “Ouest Industries Cr\'eatives” aims to strengthen Research, Education \& Innovation of the Region Pays de Loire in the field of Creative Industries. He is mostly engaged in research dealing with cognitive computing and the application of human vision modeling in image and video processing. His current centers of interest are AI boosted QoE Quality of Experience assessment, Visual Attention modeling and applications. He is co-author of more than 300 publications and communications and co-inventor of 16 international patents on these topics. He serves or has been served as associate editor or guest editor for several Journals such as IEEE TIP, IEEE STSP, IEEE TCSVT, Springer EURASIP Journal on Image and Video Processing, and SPIE JEI. He is serving in IEEE IVMSP-TC (2015- to present) and IEEE MMSP-TC (2015-to present) and one the founding member of EURASIP TAC (Technical Areas Committee) on Visual Information Processing.
\end{IEEEbiography}

\begin{IEEEbiography}[{\includegraphics[width=1in,height=1.25in,clip,keepaspectratio]{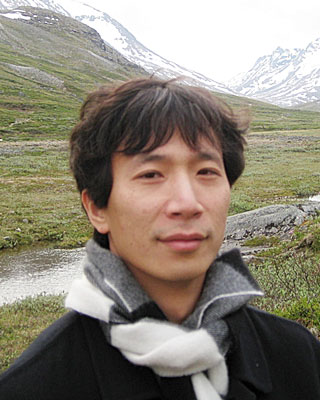}}]{Jong-Seok Lee}
(M06-SM14) received his Ph.D. degree in electrical engineering and computer science in 2006 from KAIST, Korea. From 2008 to 2011, he worked as a researcher at Swiss Federal Institute of Technology in Lausanne (EPFL), Switzerland. Currently, he is an associate professor in the School of Integrated Technology at Yonsei University, Korea. His research interests include multimedia signal processing and machine learning. He is an author or co-author of over 150 publications. He serves as an editor for the IEEE Communications Magazine and Signal Processing: Image Communication.

\end{IEEEbiography}

\end{document}